\documentclass{aa}

\usepackage{natbib}
\usepackage{threeparttable}
\usepackage{graphicx,dcolumn}
\usepackage{adjustbox,booktabs}
\usepackage[version=4]{mhchem}
\usepackage{amsmath,amssymb,latexsym}
\usepackage[varg]{txfonts}

\newcolumntype{d}{D{.}{.}{3.2}}

\begin{document}

\title{Photoinduced polycyclic aromatic hydrocarbon dehydrogenation:}
\subtitle{Molecular hydrogen formation in dense PDRs}
 
\author{P.~Castellanos\inst{1,2} \and A.~Candian\inst{1} \and H.~Andrews\inst{3} \and A.~G.~G.~M.~Tielens\inst{1}}

\date{Received 12 April, 2018 / Accepted 03 June, 2018}

\titlerunning{PAH dehydrogenation}
\authorrunning{Castellanos et al.}

\institute{Leiden Observatory, Leiden University, P.O. Box 9513, NL-2300 RA Leiden, The Netherlands\\ \email{pablo@strw.leidenuniv.nl} \and Sackler Laboratory for Astrophysics, Leiden Observatory, Leiden University, P.O. Box 9513, NL-2300 RA Leiden, The Netherlands \and Faculty of Aerospace Engineering, Delft University of Technology, Kluyverweg 1, NL-2629 HS Delft}

\abstract{The physical and chemical conditions in photodissociation regions (PDRs) are largely determined by the influence of far ultraviolet radiation. Far-UV photons can efficiently dissociate molecular hydrogen, a process that must be balanced at the H{\sc i}/\ce{H2} interface of the PDR. Given that reactions involving hydrogen atoms in the gas phase are highly inefficient under interstellar conditions, \ce{H2} formation models mostly rely on catalytic reactions on the surface of dust grains. Additionally, molecular hydrogen formation in polycyclic aromatic hydrocarbons (PAHs) through the Eley-Rideal mechanism has been considered as well, although it has been found to have low efficiency in PDR fronts. In a previous work, we have described the possibility of efficient \ce{H2} release from medium to large sized PAHs upon photodissociation, with the exact branching between H-/\ce{H2}-loss reactions being molecule dependent. Here we investigate the astrophysical relevance of this process, by using a model for the photofragmentation of PAHs under interstellar conditions. We focus on three PAHs cations (coronene, ovalene and circumcoronene), which represent three possibilities in the branching of atomic and molecular hydrogen losses. We find that, for ovalene (\ce{H2}-loss dominated) the rate coefficient for \ce{H2} formation reaches values of the same order as \ce{H2} formation in dust grains. This result suggests that this hitherto disregarded mechanism can account, at least partly, for the high level of molecular hydrogen formation in dense PDRs.}

\keywords{astrochemistry -- ISM: molecules -- molecular processes -- photon-dominated region (PDR)}

\maketitle

\section{Introduction}
\label{sec:intro}

Photodissociation regions (PDRs) correspond to mainly neutral areas in the interstellar medium (ISM) where the physical and chemical conditions are to a large extent determined by the UV field \citep[see][]{hol97}. These include a variety of environments, from the neutral gas surrounding luminous H{\sc ii} regions , reflection nebulae, the galactic warm neutral medium and diffuse clouds, among others. As their name indicates, PDRs are regions where the transition from neutral, atomic gas to its molecular counterpart takes places \citep{ste14,bia16}. As such, the balance between formation and photodissociation of the simplest, diatomic molecules is of special interest, particularly considering that chemical reactions in the gas phase are highly unlikely. Given that \ce{H2} is the most abundant interstellar molecule, its formation mechanism is of particular importance \citep[see][for a recent review]{Wakelam2017}. Most works on the formation of molecular hydrogen focus on two mechanisms. On the one hand, there are Langmuir-Hinshelwood reactions, where two physisorbed H atoms migrate on a dust grain surface and, after reacting, are released as \ce{H2} \citep[e.g.,][]{pir97,kat99}. On the other hand, we have Eley-Rideal mechanisms, where a gas phase hydrogen atom reacts with another, chemisorbed hydrogen and both are desorbed as \ce{H2} \citep[e.g.,][]{dul96,hab04}. The former process is known to be efficient at low dust temperatures ($T_\mathrm{d} \leq 20$~K), while the latter can be effective at higher temperatures, more representative of dense PDRs.

Polycyclic aromatic hydrocarbons (PAHs) emission dominates the mid-infrared spectrum of a wide variety of astronomical sources containing dust and gas, including PDRs \citep[and reference therein]{Tielens2013}. The peak emission of PAH bands is found to coincide with \ce{H2} emission lines, prompting the suggestion of a link between both species \citep{hab03}. This link can be due to \ce{H2} formation via \ce{H2} abstraction from superhydrogenated PAHs \citep{rau08} or the \ce{H2}-loss from the fragmentation of PAHs \citep{joc94}. The latter mechanism is associated with the general idea of interstellar top-down chemistry with PAHs as precursors. Observations of PDRs have revealed that, as the PAH band intensity decreases, the fullerene (\ce{C60}) band intensity increases \citep{ber12,cas14}. This suggests that photoprocessing of PAHs leads to their fragmentation and eventual formation of fullerenes among other intermediate species. Furthermore, several small hydrocarbons --- such as \ce{C2H}, \ce{C3H2}, \ce{C3H+} and \ce{C4H} --- are known to be related to PAH mid-IR emission bands in PDRs, an association that has also been considered indicative of a top-down formation mechanism \citep{pet05,guz15,cua15}. The formation of small and large hydrocarbon products through top-down photochemistry requires intense UV-fields, which is consistent with the observations of the aforementioned molecules deep within the atomic part of th PDR. However, \ce{H2} formation will be effective in slightly more shielded environments, namely the H{\sc i}/\ce{H2} transition region.

\begin{figure*}
  \centering
  \includegraphics[width=\textwidth]{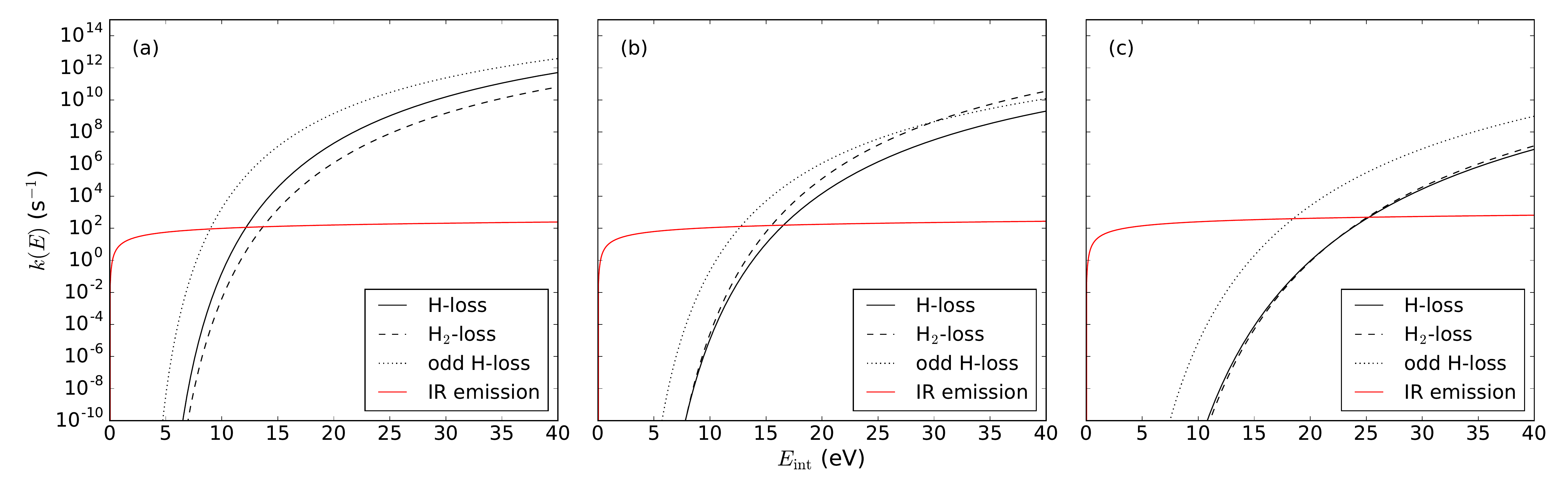}
  \caption{Hydrogen loss rates as a function of internal energy for the coronene (a), ovalene (b) and circumcoronene (c) cations. H- and \ce{H2}-loss are considered as from the parent molecule, while odd H-loss refers to hydrogen loss from a partly filled duo.}
  \label{fig:ratesE}
\end{figure*}

\citet{hab04} used the ratio of rotational to rovibrational \ce{H2} lines to determine the formation rate coefficient ($R(\mathrm{H_2})$) in dense, hot PDRs. Their estimate yields values of $R(\mathrm{H_2}) \sim 10^{-16}$~cm$^3$~s$^{-1}$, almost an order of magnitude higher than the standard value found in diffuse clouds, where \ce{H2} formation is expected to come exclusively from dust grains through an Eley-Rideal mechanism \citep[$3\times 10^{-17}$~cm$^3$~s$^{-1}$][]{jur75,gry02}. The \ce{H2} formation rate in the diffuse ISM is generally well accounted for when considering only grain surface reactions \citep{hol71}, but these reactions alone fall short at explaining the situation in dense PDRs. In order to account for this deficiency, reactions involving PAHs have been invoked as an additional mechanism leading to \ce{H2} formation. Such a connection is further supported, seeing as there is a clear spatial correlation linking \ce{H2} and PAH emission \citep{hab03}. The proposed routes include an Eley-Rideal mechanism analogous to that described for dust, involving the abstraction of molecular hydrogen from superhydrogenated PAHs \citep{rau08,men12}, and direct photodissociation of PAHs involving \ce{H2}-loss \citep{joc94,Ling95}. The importance of both formation pathways has been investigated using different models for a variety of PDR conditions \citep[e.g.,][]{bro14,Bosch15,andrews16}. These models have established that surface reactions (i.e., Eley-Rideal) on PDR fronts cannot efficiently form \ce{H2}, given that under these conditions superhydrogenated PAHs make up a nearly negligible fraction of the total population. While photodissociation through \ce{H2}-loss is found to be the dominant \ce{H2} formation pathway from PAHs, its rate makes a negligible contribution to the total \ce{H2} formation in PDRs. However, we note that for \ce{H2}-loss from PAHs, all models rely on photodissociation parameters extrapolated from experimental and theoretical work on rather small PAHs (up to 24 C-atoms).

On a previous work modeling the experimental dissociation pattern of PAHs, we have determined the barriers and enthalpy change required for H- and \ce{H2}-loss from medium sized PAHs \citep{cas18}. That study included an analysis on small, previously studied PAHs in order to validate our results against previous work. We found that H-hopping within PAHs is an essential part in the dehydrogenation process, creating an aliphatic-like side group (\ce{CH2}), which can lead to the release of atomic or molecular hydrogen. In small PAHs, H-loss is found to be the dominant channel, in accordance with other results found in the literature \citep{joc94,Ling95,Ling98}. Large PAHs show the predominant channel to be \ce{H2}-loss, although with the caveat that solo hydrogens can only be lost in atomic form from aromatic positions. 

In the present study we utilize our previously calculated rates for coronene and ovalene, and expand our calculations to circumcoronene. In Sect.~\ref{sec:methods} we discuss our model for \ce{H2} formation through photolysis of PAHs. The results are presented in Sect.~\ref{sec:results} and in Sect.~\ref{sec:conclusions} we discuss \ce{H2} formation in PDRs and summarize our conclusions.

\section{Methods}
\label{sec:methods}

Density Functional Theory (DFT) calculations on the structures of intermediate and transition states were performed using B3LYP/6-31G(d,p) and the quantum chemistry software Gaussian~09 \citep{fri09}. Transition state structures were in most cases determined with the Berny algorithm, while the Synchronous Transit-Guided Quasi-Newton (STQN) method \citep{STQN1,STQN2} was used in more complex cases. Reaction rates were calculated using Rice-Ramsperger-Kassel-Marcus (RRKM) theory \citep{baer-hase}, using the sum of states for the transition state and the density of states for the parent structures. The sum and density of states were calculated using the normal vibrational modes derived from DFT, by means of the \texttt{densum} program from the MULTIWELL suite \citep{Multiwell1,Multiwell2}.

\begin{figure*}
  \centering
  \includegraphics[width=\textwidth]{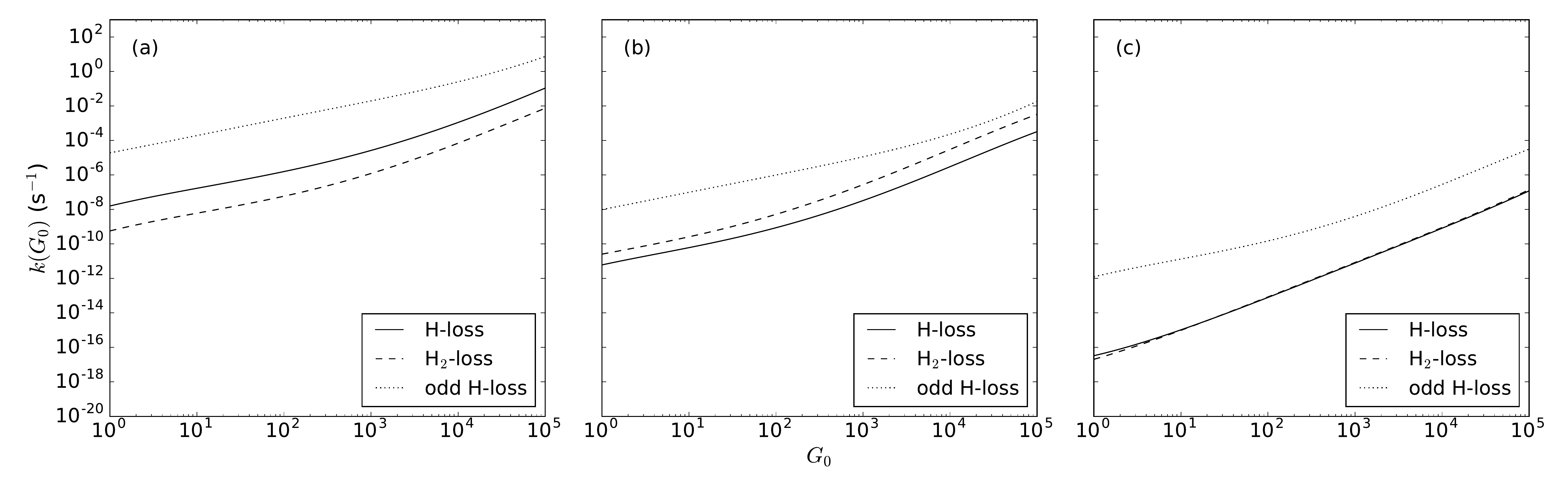}
  \caption{Hydrogen loss rates as a function of $G_0$ for the coronene (a), ovalene (b) and circumcoronene (c) cations. H- and \ce{H2}-loss are considered as from the parent molecule, while odd H-loss refers to the loss from a partly filled duo.}
  \label{fig:ratesG0}
\end{figure*}

The aforementioned methods were used to study the dehydrogenation of a series of PAHs, together with experimental and Monte-Carlo techniques \citep{cas18}. The results of that work show that the fragmentation process is dominated by formation of ``aliphatic'' groups due to hydrogen roaming --- i.e., the formation of \ce{CH2} groups on the edge of the molecule. From this point on, dehydrogenation can proceed either by atomic or molecular hydrogen loss, in a manner that is molecule dependent. In this work, we will use the results already derived for coronene and ovalene. As mentioned in \citet{cas18}, in order to fit the experimental data of ovalene, the energy barriers for aliphatic H- and \ce{H2}-loss had to be adjusted to match those of coronene, while keeping the values for $\Delta S_{1000}$ as derived from the DFT calculations. The original study did not include the circumcoronene cation and experiments on the dehydrogenation of this molecule are currently lacking. However, we have calculated the intermediate structures and transition states involved in the formation of aliphatic groups and the loss of atomic and molecular hydrogen. In order to keep consistency with the previous calculations, the energy barriers for the loss channels in circumcoronene have also been modified to match those of coronene.

We note that coronene, ovalene and circumcoronene all have duo hydrogens and, in the case of ovalene and circumcoronene, solo hydrogens with no bay-regions. Following the conclusions from \citet{cas18}, we have disregarded reactions involving tertiary carbons (an edge carbon bridging two neighboring rings). Aliphatic formation is thus only possible within duos, and aromatic H-loss is the only fragmentation channel available for the solos. The isomerization rates considered in this work involve the rate of formation of an aliphatic group and its reverse. For photodissociation, both aliphatic and aromatic H- and \ce{H2}-losses are calculated. All rates are based on the fully hydrogenated form of the PAH in question, with the exception of aromatic H-loss from a partially hydrogenated duo, for which the parent structure is that of the fully aromatic PAH after a single H-loss. The latter loss rate is considered separately, given that the energy barrier is $\sim$1~eV lower than for a full duo.

\begin{figure*}
  \includegraphics[width=\textwidth]{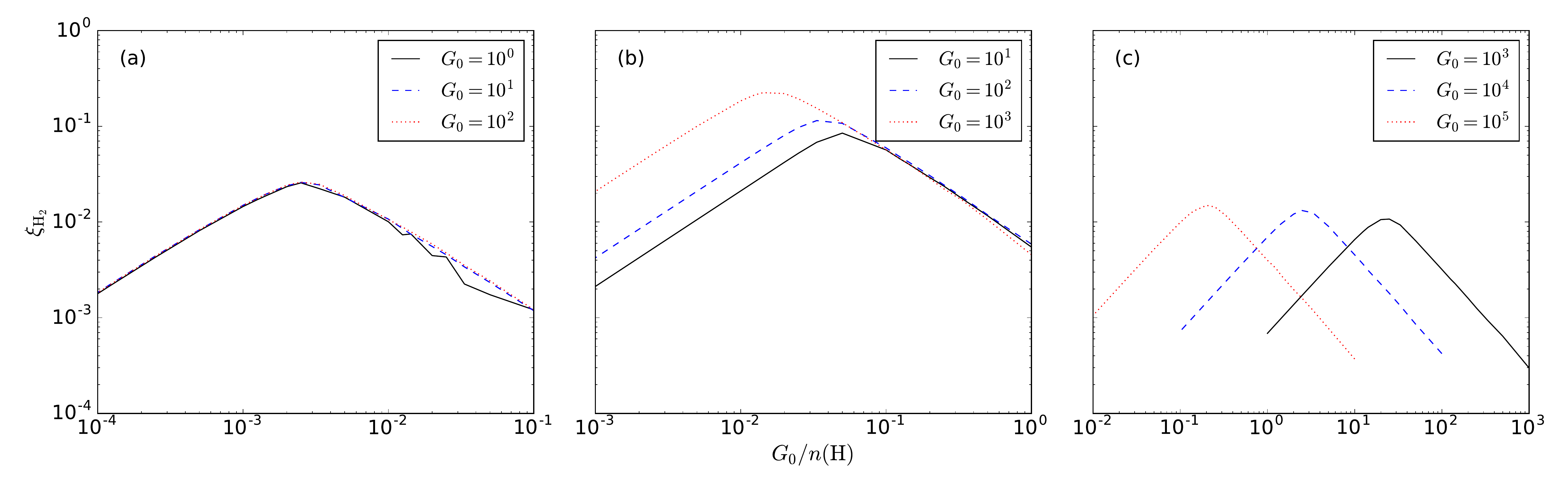}
  \caption{\ce{H2} formation efficiency in the photo-fragmentation of the coronene (a), ovalene (b) and circumcoronene (c) cations in a variety of interstellar environments.} 
  \label{fig:eff}
\end{figure*}

In order to assess the astronomical relevance of \ce{H2} formation, we implemented our calculated rates into the model for hydrogenation and dehydrogenation of PAHs of \citet{andrews16}. Their model includes a variety of physical and chemical processes for PAHs in interstellar environments, including ionization, electron recombination, hydrogen attachment, photodissociation (as H- and \ce{H2}-loss) and \ce{H2} formation from superhydrogenated molecules following Eley-Rideal abstraction. The focus of the present work is to investigate \ce{H2} formation through photodissociation on PDRs. Given that the transition from fully hydrogenated PAHs to fully dehydrogenated PAHs (the PDR front) occurs at fairly high values of UV-field intensity ($G_0$), processes involving superhydrogenated PAHs --- such as Eley-Rideal abstraction --- can be disregarded. Under the same considerations, and taking into account that our DFT calculations are focused on cations, we keep electron recombination to a minimum by setting $n_e = 10^{-7}$~cm$^{-3}$, thus ensuring that all PAHs in the model are singly ionized. While this value does not reflect the true electron density in PDRs, the efficiency of \ce{H2} formation from PAH cations is not affected by $n_e$.

The photodissociation rates considered in the model by \citet{andrews16} are considered in terms of H- and \ce{H2}-loss. Given that in their original formulation the formation of aliphatic groups via hydrogen hopping was not taken into account, the model receives as input a single rate for each loss. In order to work around this issue, we have calculated effective photodissociation rates. For this, the fraction of the time the PAH in question exists with an aliphatic group is calculated using
\begin{equation}
  f_\mathrm{al}(E) = \frac{k_\mathrm{al}(E)}{k_\mathrm{al}(E)+k_\mathrm{al,r}(E)},
\end{equation}
where $k_\mathrm{al}(E)$ is the aliphatic formation rate as a function of energy, and $k_\mathrm{al,r}(E)$ is the rate for the reverse reaction. For both rates the corresponding degeneracies are taken into account --- e.g.\ in fully hydrogenated coronene, the twelve hydrogen atoms can hop towards the neighboring H in the same duo, while for the reverse reaction either of the two hydrogen atoms can hop back into the empty carbon. The time fraction the molecule is in a fully aromatic configuration, on the other hand, will be given by $1-f_\mathrm{al}(E)$. The effective H-loss rate is calculated by multiplying the respective fractions to the corresponding photodissociation rates and adding together the contributions of aliphatic and aromatic losses. The same process is applied to \ce{H2}-loss.

\section{Results}
\label{sec:results}

The effective fragmentation rates, as derived from DFT calculations, are shown in Fig.~\ref{fig:ratesE}. The three PAHs selected for this work provide examples of the three types of photofragmentation balance that can be found: H-loss dominated (coronene), \ce{H2}-loss dominated (ovalene) and competitive (circumcoronene). While these rates include contributions from aromatic rates, aliphatic losses are in all cases the dominant channel for the respective loss by as much as five orders of magnitude in the relevant energy range. The only exception to this is the odd hydrogen loss, for which the aromatic H-loss dominates in all three molecules. For fragmentation to take place, the dominant loss rate must exceed the IR photon emission rate, which is the case at 12.2, 15.5 and 25.1~eV for coronene, ovalene and circumcoronene, respectively. Considering the 13.6~eV cutoff for UV photons in the ISM, this means that, while coronene can be fragmented by single photon absorption, PAH dehydrogenation will be dominated by multiple photon absorption.

Figure~\ref{fig:ratesG0} shows the fragmentation rates as a function of $G_0$. These were calculated following the procedure described by \citet{andrews16}, taking into account the temperature probability function following \citet[and references therein]{bak01}. Such a model considers as well the possibility of multiphoton events which, as mentioned, are needed in order to fragment larger PAHs. The effects of multiple photon absorption can be observed in Fig.~\ref{fig:ratesG0}, as deviations from a linear behavior are observed at $G_0 = 10^3$ in coronene, but earlier on for ovalene and circumcoronene. The three variations on the H-/\ce{H2}-loss balance described before are still evident here, with odd H-losses dominating in cases where it is available. The effects of degeneracy are minor in all three PAHs, with variations due to hydrogenation level remaining below an order of magnitude in all cases.

The fragmentation of the PAHs studied here follows an established pattern under interstellar conditions. When hydrogen within duos are part of the molecule in question, these are the preferred dehydrogenation sites in all cases. Even hydrogen losses can proceed via aliphatic H- or \ce{H2}-loss according to the characteristics of the PAH under consideration, as explained before. Odd hydrogen losses always occur via aromatic C--H bond cleavage with a reduced barrier. Considering that the rates involved in H-loss from the solos are exclusively aromatic and do not have the reduced barrier of partly hydrogenated duos, these are the last hydrogens to be lost in the molecule.

We have calculated the efficiency of \ce{H2} formation ($\xi_\mathrm{H_2}$) considering the fragmentation and hydrogen addition rates as a function of the hydrogenation state ($i$) using,
\begin{equation}
 \xi_\mathrm{H_2}(Z) = \sum_i\frac{2k_\mathrm{H_2}(Z,i)f(Z,i)}{\kappa_\mathrm{add}n(\mathrm{H})},
\end{equation}
where $n(\mathrm{H})$ corresponds to the atomic hydrogen density, $\kappa_\mathrm{add}$ is the hydrogen addition rate coefficient and $f(i)$ corresponds to the fraction of the PAH with hydrogenation state $i$. For cations, we use $\kappa_\mathrm{add} = 1.4\times 10^{-10}$, following \citet{andrews16}. Given that in our model we only take into account a single ionization state, the following results represent the total \ce{H2} formation efficiency, based on the cations. Figure~\ref{fig:eff} presents the efficiency of coronene, ovalene and circumcoronene under different UV field intensities and atomic hydrogen densities. Coronene and circumcoronene exhibit $\xi_\mathrm{H_2}$ in the order of 1--2\% under all the environmental conditions explored here. More interestingly, ovalene shows efficiencies at least one order of magnitude higher, depending on the values of $G_0$ and $n(\mathrm{H})$. However, \ce{H2} formation efficiency in ovalene does not approach one at any $G_0$, even though one could naively expect this given that aliphatic losses are heavily dominated by \ce{H2}-loss. The source of this apparent contradiction is due to the highly efficient loss from partly hydrogenated duos. Hydrogen addition cannot efficiently compete with this form of dehydrogenation, quenching the full rehydrogenation of duos which in turn prevent the formation of a new \ce{H2} unit.

In the case of coronene, the efficiency exhibits the same behavior as a function of $G_0$, independently of $G_0$. This is due to fact that coronene fragmentation at these values of $G_0$ is dominated by single photon events (Fig.~\ref{fig:ratesG0}). Ovalene and, more prominently, circumcoronene show a large spread in the peak position of $\xi_\mathrm{H_2}$ with respect to $G_0/n(\mathrm{H})$ as a function of $G_0$. This again demonstrates the degree of non-linearity of multiphoton events in PAH fragmentation. From the results presented in Fig.~\ref{fig:ratesE}, it is evident that two-photon events will dominate the fragmentation of ovalene while, for circumcoronene, three photons are required to initiate the dehydrogenation process.

An increase in the peak value of $\xi_\mathrm{H_2}$ with $G_0$ reflects the result of the competition between the dehydrogenation channels. Ovalene shows an evident increase in the peak value of the efficiency of molecular hydrogen formation, reflecting the increase of $k_\mathrm{H_2}$ with increasing $G_0$ and the near lack of competition with H-loss channels. In coronene, on the other hand, no such increase in $\xi_\mathrm{H_2}$ is observed, as it is to be expected considering that H-loss is the dominant channel. Unlike coronene, circumcoronene does show a slight increase in \ce{H2} formation efficiency as $G_0$ increases, but it is nowhere near as evident as for ovalene. The fact that molecular and atomic hydrogen loss are in close competition prevents the increase of $\xi_\mathrm{H_2}$ to the extent observed in ovalene.

\begin{figure*}
  \includegraphics[width=\textwidth]{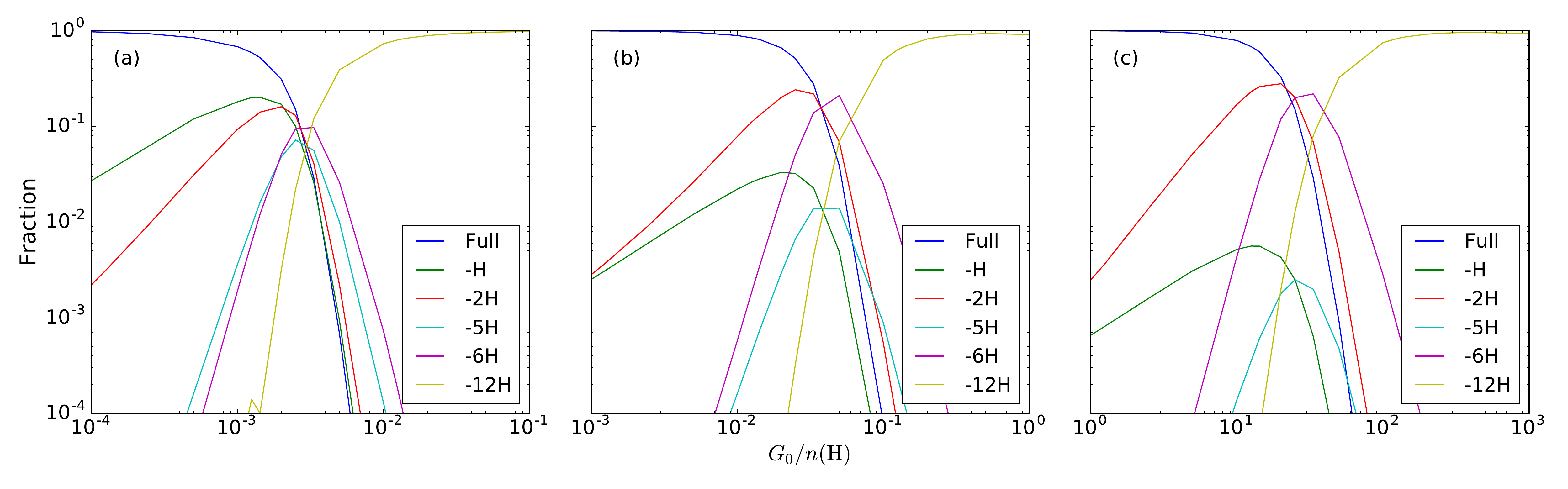}
  \caption{Fragmentation pattern for the coronene (a), ovalene (b) and circumcoronene (c) cations. They are taken at 10, 100 and 1000~$G_0$, respectively. -12H corresponds to the point in all molecules where aliphatic loss is no longer possible. Solo hydrogens are still left in the molecule at this stage in ovalene and circumcoronene (two and six, respectively).} 
  \label{fig:hydro}
\end{figure*}

Figure~\ref{fig:hydro} shows the contribution of different hydrogenation states to the total population of the molecules, at selected values of $G_0$. The peak in the corresponding efficiency curves (Fig.~\ref{fig:eff}) coincides with the value of $G_0/n(\mathrm{H})$ that has the largest contribution from the -6H fragment in all cases. This is to be expected, given that dehydrogenation from duos is the only pathway contributing to \ce{H2} formation and coronene, ovalene and circumcoronene have the same number of duos. For PAHs with a different number of duos or other types of edge structures this situation will vary, but in general molecular hydrogen formation efficiency peaks at the point where half of the structures capable of producing \ce{H2} have become dehydrogenated.

The dehydrogenation pattern shows an enhancement of even hydrogen molecules with respect to odd ones, a well known experimental fact \citep{cas18}. However, the degree to which PAH fragments with an even number hydrogens dominates depends on a number of factors. Most notoriously, the dominant fragmentation channel plays a significant role. When comparing coronene and ovalene (Figs.~\ref{fig:hydro}a and b) it is clear that the fraction of fragments with an odd number of hydrogens in coronene has a larger contribution than that of ovalene. This is to be expected given that fragmentation from coronene is dominated by H-loss. However, there is also a dependency on how dominant the odd H-loss is over the even dehydrogenation channels. Case in point, circumcoronene has higher contributions from photoproducts with an even number of hydrogens than ovalene, although the latter is clearly dominated by \ce{H2}-loss dominated, while the former shows a strong competition. Nevertheless, the odd H-loss rate in circumcoronene is faster than any of the even hydrogen loss channels, which leads to a reduced contribution of fragments with an odd hydrogenation.

The tail of $\xi_\mathrm{H_2}$ at large $G_0/n(\mathrm{H})$ must be considered carefully. Once full dehydrogenation is present, the carbon skeleton can isomerize into cages and/or further fragment \citep{ber12,zhe14b}. This will remove the carbon ``flake'' from the population, rendering meaningless the calculated value of $\xi_\mathrm{H_2}$. However, the presence of solo hydrogens appears to act as a bottleneck in the fragmentation process. As seen in Fig.~\ref{fig:hydro}, ovalene and circumcoronene retain their last few hydrogens even at fairly large values of $G_0/n(\mathrm{H})$. Nevertheless, one must take into account that this model does not consider isomerizations of the carbon skeleton, neither before nor after full dehydrogenation has taken place. Given that the rate of aromatic H-loss from solo hydrogens is rather slow, the internal energy of the molecule can build up further than when aliphatic losses are available. This can, in turn lead to other isomerization routes and/or fragmentation via C containing units. 
 
\section{Discussion and Conclusions}
\label{sec:conclusions}

Using the results of the previous section, we can give a simplified expression for $\xi_\mathrm{H_2}$. From Fig.~\ref{fig:hydro} it is evident that, while the dehydrogenation process is ongoing, most of the molecules will have an even hydrogenation level. This, together with the fact that variations of $k_\mathrm{H_2}$ with dehydrogenation are not very significant, leads to,
\begin{equation}
  \xi_\mathrm{H_2} = \frac{2k_\mathrm{H_2}}{\kappa_\mathrm{add}n(\mathrm{H})}.
\end{equation}
This approximation is generally valid within a factor two up to the peak of the molecular hydrogen formation efficiency and will rapidly diverge from the exact value at large $G_0/n(\mathrm{H})$. The values of $G_0$ and $n(\mathrm{H})$ at which $\xi_\mathrm{H_2}$ will peak are, unfortunately, more involved to calculate and a simple analytic expression valid for all PAHs cannot be derived. This is due to the effect of multiphoton absorption (a non-linear process) being an indispensable part of PAH dehydrogenation. Additionally, the dissociation parameters determined here show a wide range of variations among different PAHs, even within the same ``family'', as seen for coronene and circumcoronene. The unpredictability of the peak position of \ce{H2} formation efficiency showcases how critical it is to derive accurate photodissociation parameters (in the form of energy barriers and activation enthalpies) and to have experimental confirmation of such values. 

The values derived here for $\xi_\mathrm{H_2}$ at high values of $G_0/n(\mathrm{H})$ have to be carefully considered. Fully dehydrogenated PAHs will rapidly undergo isomerization into fullerenes and other compounds that are not accounted for here. While the presence of solo hydrogens in PAHs appears to delay the onset of the full dehydrogenation, the high value of internal energy that the molecule can reach opens up the possibility of other unimolecular reactions not considered thus far. This caveat should not erode the fact that, in regions where partial hydrogenation makes up the larger fraction of the PAH population, the values derived here are highly accurate.

Even though the exact extent of \ce{H2} formation through PAH fragmentation is heavily dependent on the specifics of the molecule under investigation, the current results do paint a more positive picture of the process than previous work. In a manner analogous to the formation rate in dust grains \citep[$R_\mathrm{d}$][]{hol79,bur83}, the \ce{H2} formation rate in PAHs has been defined by \citet{andrews16} as,
\begin{equation}
  R_\mathrm{PAH}(\mathrm{H_2}) = \frac{1}{2}\xi_\mathrm{H_2}\kappa_\mathrm{add} X_\mathrm{PAH} \simeq 4\times 10^{-18} \left(\frac{f_\mathrm{C}}{0.1}\right)\left(\frac{50}{N_\mathrm{C}}\right)\left(\frac{\xi_\mathrm{H_2}}{0.1}\right)~\mathrm{cm^3s^{-1}},
\end{equation}
where $X_\mathrm{PAH}$ corresponds to the abundance of PAHs per hydrogen atom, $f_\mathrm{C}$ is the fraction of elemental carbon locked in PAHs and $N_\mathrm{C}$ is the number of C-atoms per PAH. Since the most favorable case within our PAH sample corresponds to ovalene, we will be basing our calculation on its results and using the physical conditions for the well studied NW PDR of NGC~7023. Following this we find $R_\mathrm{PAH}(\mathrm{H_2}) \sim 10^{-17}$~cm$^3$s$^{-1}$. This is comparable to the \ce{H2} formation rate on dust grains --- $1.5\times 10^{-17}$~cm$^3$~s$^{-1}$ at $T_\mathrm{gas} = 400$~K and $T_\mathrm{d} =35$~K --- as calculated from the standard expression given by \citet{hol79}. While the value for $R_\mathrm{PAH}(\mathrm{H_2})$ calculated here is about two orders of magnitude higher than the previous estimate \citep{andrews16}, it is nevertheless not enough to account for the disparity between the \ce{H2} formation rate in PDRs as derived from calculations and observations \citep{hab04}. As was recognized by \citet{andrews16}, Eley-Rideal abstraction from superhydrogenated PAHs is not an efficient process within the photodissociation front of NGC~7023, where they note \citet{bro14} use a value two order of magnitude higher than the experimentally determined cross-section for coronene films \citep{men12}.

Molecular hydrogen formation from the photo-fragmentation of PAH cations can be an important process in PDRs, particularly in the region of transition between H{\sc i} and \ce{H2}. Although the \ce{H2} formation rate derived here is still unable to fully account for the large, observationally determined values in dense PDRs, they represent a step forward towards bridging the gap. Furthermore, these results lend further support to the case for top-down interstellar chemistry, with PAHs as a starting point, to explain the presence of molecules which are otherwise hard to account for using more traditional chemical pathways. The exact contribution to the overall \ce{H2} formation is heavily dependent on molecular properties in ways that have yet to be fully understood, and contributions from PAHs in other ionization states must be considered as well. A solution to this issue lies in close collaborations of theoretical and experimental work in order to determine the precise dissociation parameters. Nevertheless, it is evident from this work that, in molecules were the dominant dissociation channel is \ce{H2}-loss, $\xi_\mathrm{H_2}$ far exceeds previous estimates. Under such conditions, molecular hydrogen formation from PAHs can be on par --- and even exceed --- the estimated contribution of molecular hydrogen formation from dust grains through Eley-Rideal \ce{H2} abstraction.

\begin{acknowledgements}
Studies of interstellar chemistry at Leiden Observatory are supported through advanced-ERC grant 246976 from the European Research Council, through a grant by the Netherlands Organization for Scientific Research (NWO) as part of the Dutch Astrochemistry Network, and a Spinoza premie. We acknowledge the European Union (EU) and Horizon 2020 funding awarded under the Marie Sk\l{}odowska-Curie action to the EUROPAH consortium, grant number 722346. AC acknowledges NWO for a VENI grant (number 639.041.543). DFT calculations were carried out on the Dutch national e-infrastructure (Cartesius) with the support of SURF Cooperative, under NWO EW projects MP-270-13 and SH-362-15.
\end{acknowledgements}

\bibliographystyle{aa}
\bibliography{references}

\end{document}